\documentclass[11pt]{article}

\usepackage{amsmath}
\usepackage{amssymb}
\usepackage{times}
\usepackage{dsfont}
\usepackage{theorem}
\usepackage{graphics} 
\usepackage{cancel}
\usepackage{graphicx}
\usepackage[section]{placeins}
\usepackage{hyperref}

\title{Point kinetic model of the early phase\\
of a spherically symmetric nuclear explosion}
\author{Andreas Walter Aste\\
$\quad$\\
\emph{Department of Physics, University of Basel, 4056 Basel, Switzerland}}

\date{June 6, 2016}

\begin{document}
\maketitle

\begin{abstract}
A concise point kinetic model of the explosion of a prompt supercritical sphere driven by a nuclear fission chain reaction
is presented. The findings are in good agreement with the data available for Trinity,
the first detonation of a nuclear weapon conducted by the United States Army as part of the Manhattan project.
Results are presented for an implosion device containing pure plutonium-239, although the model can be easily applied to, e.g.,
uranium-235. The fizzle probability and corresponding yield of a fission bomb containing plutonium recovered from reactor
fuel and therefore containing significant amounts of spontaneously fissioning plutonium-240 which can induce
a predetonation of the device is illustrated by adding a corresponding source term in the presented model.
Related questions whether a bomb could be made by developing countries or terrorist organizations can be tackled this way.
Although the information needed to answer such questions is in the public domain, it is difficult to extract a consistent
picture of the subject for members of organizations who are concerned about the proliferation of nuclear
explosives. 
\\
\vskip 0.1 cm {\bf Physics and Astronomy Classification Scheme (2010).} 28.20.-v Neutron physics; 25.85.Ec Neutron-induced fission;
28.70.+y Nuclear explosions.\\
\end{abstract}

\FloatBarrier
\section{Introduction}
Despite the terrifying fact that numerous operational fission or even thermonuclear bombs exist on our planet, there is a great interest
in the basic principles and physics underlying the concept of nuclear weapons. During the last decades, an increasing number of details
concerning the design of the Gadget, the first atomic bomb ignited in the Trinity test, have been revealed.
In the present work, the first man-made nuclear explosion is modeled in a simplified but instructive manner, which nevertheless
leads to realistic and even rather accurate quantitative results. The calculations are based on a point model in the analogous sense of nuclear
reactor point models or neutron point kinetics, where the whole structure of the reactor core is averaged out in an effective way,
removing the spatial structure of the object under study which  therefore becomes a structureless 'point', but still retaining the basic physical
features in the time domain. 

\section{Point kinetic model}
\begin{figure}[!htbp]
\begin{center}
\includegraphics[width=6.2cm,angle=270]{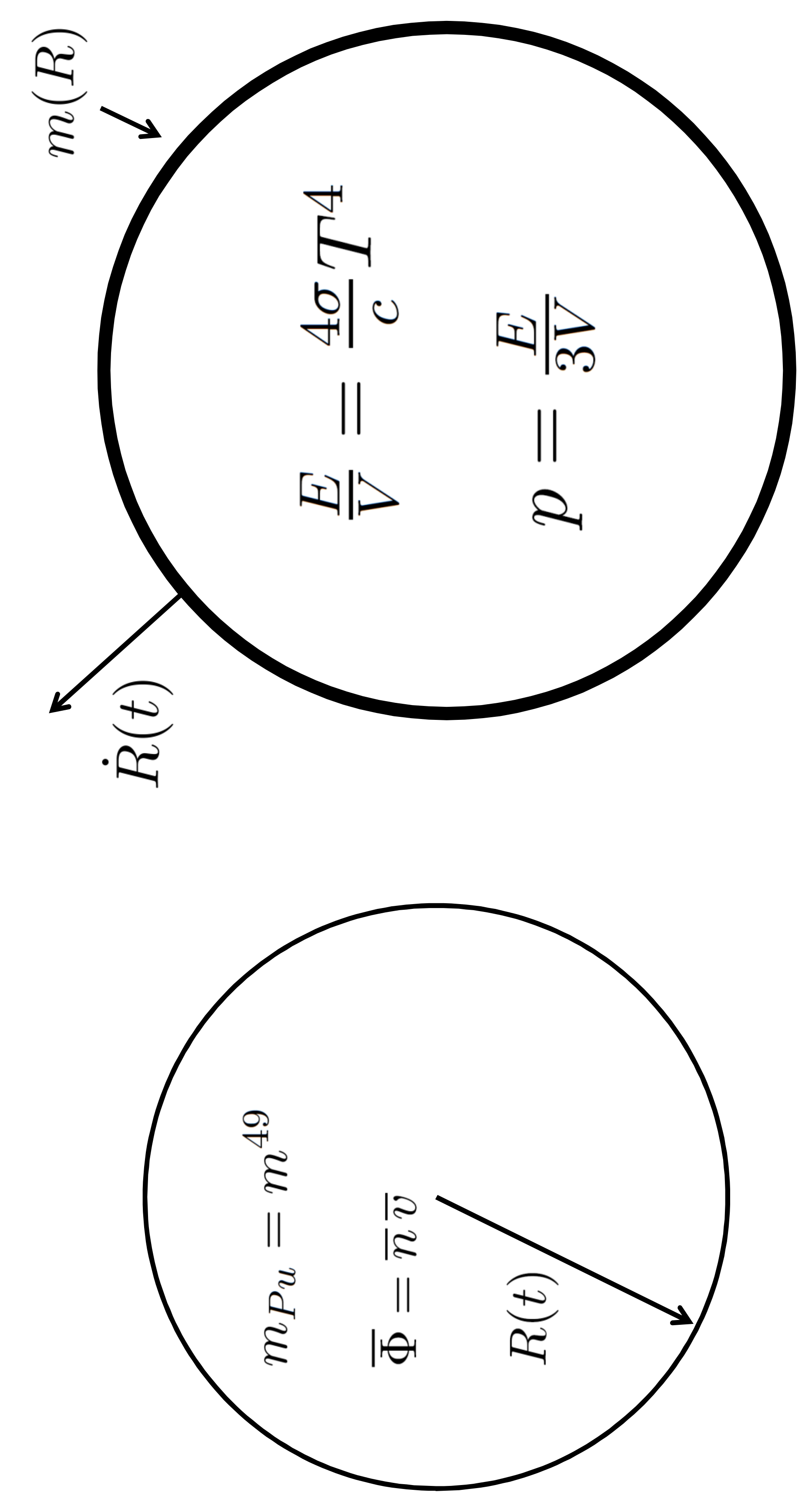}
\caption{Basic assumptions of the point kinetic model: (A) The matter of the plutonium core moves on a sphere with velocity
$\dot{R} (t)$, including all the matter originally located inside the sphere with radius $R(t)$. (B) The expansion of the core is
driven by the radiation pressure inside the matter-free sphere, which converts the energy released from fission to the matter's
kinetic energy. (C) The dynamics of the chain reaction is inspired from ordinary neutron kinetics and transport theory inside a homogeneous
plutonium sphere.}
\label{fig_model}
\end{center}
\end{figure}

Below, we will model a exploding plutonium core including its surrounding matter both as a sphere or a ball with a time-dependent radius $R(t)$,
where the expansion is driven by (a gradient of) the radiation pressure produced by the energy released by the nuclear chain reaction
(see also Fig. (\ref{fig_model})). On the one hand, the time-dependent density of fissile $^{23{\bf{9}}}_{\, \, 9{\bf{4}}}$Pu atoms
$\rho_A^{49} (t)$ and other crucial quantities are assumed to be spatially constant inside the sphere as an approximation
and one defining feature of the present point model. On the other hand, we will assume that the explosion builds up
a fireball, a matter-, or a 'blast'- shell confined by a wall of fire in the close vicinity of the surface of the sphere of radius $R(t)$, containing all the material that
originally was located inside the sphere finally compressed to a thin layer. The interior of the fireball is basically matter-free but
filled with black body radiation with a temperature $T(t)$, i.e. with a homogeneous photon gas reaching a very high energy density
and a radiation pressure of several hundred gigabars during the nuclear explosion. This picture is certainly justified within the first microseconds
after ignition of the chain reaction by a neutron source, where the explosion generates very high temperatures of several
$10^7 \rm K$ \cite{Glasstone}, but one must keep in mind that applying concepts from equilibrium thermodynamics clearly represents
an approximation. Since the mean free path of the photons inside the dense plutonium core is of the order of $10^{-3} \rm mm$,
the major part of the energy released by fission will remain in the core during this phase.
The radiation temperature must be clearly distinguished from the lower temperature distribution inside the
expanding matter shell, whose detailed structure is of minor interest in the present exposition. However, the hot matter shell
emits hazardous intense X-rays, ultraviolet light, visible light, infrared light, and thermal radiation to the environment.

Due to its symmetries, the point kinetic model shares some similarities with cosmological models of the early universe.

 \subsection{Neutron diffusion theory}
The time-dependent diffusion equation for the prompt neutron flux density $\Phi$ or the neutron density $n$
inside the sphere containing the fissionable material ($r \le R(t))$ reads
\begin{equation}
\frac{\partial n(t,\vec{r})}{\partial t} = \frac{1}{\overline{v}}  \frac{\partial \Phi (t,\vec{r})}{\partial t}
= - \,  {\rm div} \,  (- D(t,\vec{r}) \,  {\rm grad} \,  \,  \Phi(t,\vec{r})) +
(\nu \Sigma_f(t,\vec{r}) - \Sigma_a(t,\vec{r})) \Phi(t,\vec{r}) \, , \label{diffusion_model}
\end{equation}
where the neutron density $n(t)$ is related to the neutron flux density $\Phi=n \overline{v}$ by the average fission neutron
velocity. The fast neutron spectrum is not known with high accuracy, and the neutrons inside the sphere containing the
fissionable material undergo elastic but also inelastic scattering. Below, we will work with a generally accepted value
for average neutron velocity of $\overline{v}=18 \cdot 10^3 \rm km \cdot s^{-1}$.

A diffusion model according to Eq. (\ref{diffusion_model})  is justified to some extent since the velocity of fission neutrons
is generally much bigger than the velocity of the fissionable material.
According to diffusion theory, a gradient of the neutron flux density induces a neutron current density $\vec{j}$ according to
Fick's law of diffusion
\begin{equation}
\vec{j} (t, \vec{r}) = - D(t, \vec{x}) \, {\rm grad} \,  \Phi(t, \vec{r})  =- D(t, \vec{x}) \vec{\nabla} \Phi(t, \vec{r}) \, ,
\end{equation}
where $D(t, \vec{r})$ is the diffusion constant for fast neutrons inside the material under study, i.e. $D$ is a parameter
which basically describes the local interactions of the neutrons governing how easily the neutrons can move.
The divergence of the current density $\vec{j}$ equals the local neutron leakage rate per volume due to diffusion
\begin{equation}
\frac{\partial n_{d} (t, \vec{r})}{\partial t} = - {\rm div} \, \vec{j} (t, \vec{r}) =
\vec{\nabla} [ D(t, \vec{r}) \vec{\nabla} \Phi (t, \vec{r}) ] \, .
\end{equation}

As already mentioned we presume homogeneity in the sense that the diffusion constant $D(t)$
and the total, fission, absorption, capture, or scattering macroscopic cross sections
$\Sigma_{t,f,a,c,s}(t)=\sigma_{t,f,a,c,s} \rho_A(t)$ are assumed to be spatially constant inside the sphere
where the diffusion equation will be investigated. A macroscopic cross section $\Sigma$ is the collective cross section
of all atoms per volume. This collective cross section density gets permeated by the neutron flux density $\Phi$; accordingly, the
corresponding reaction rate per volume is given by the product $\Sigma \Phi$.
Given the average number $\nu$ of (prompt) neutrons released per fission,
the local neutron production rate per volume is
\begin{equation}
\frac{\partial n_f (t)}{\partial t} = \nu \Sigma_f(t) \Phi(t) \, ,
\end{equation}
and since the material inside the core absorbs neutron at a rate per volume
\begin{equation}
\frac{\partial n_a (t)}{\partial t} = -\Sigma_a (t) \Phi (t) \, ,
\end{equation}
one finally arrives at the spatially homogenized version of the balance equation (\ref{diffusion_model})
\begin{displaymath}
\frac{\partial n(t,\vec{r})}{\partial t} = \frac{\partial n_{d} (t, \vec{r})}{\partial t} +
\frac{\partial n_{f} (t, \vec{r})}{\partial t} + \frac{\partial n_{a} (t, \vec{r})}{\partial t} \, .
\end{displaymath}
\begin{equation}
=D(t) \vec{\nabla}^2 \Phi(t, \vec{r}) + \nu \Sigma_f(t) \Phi(t, \vec{r}) - \Sigma_a(t) \Phi(t,\vec{r}) \, .
\end{equation}

Of course, quantities like the density of atoms $\rho_A (t)$
and macroscopic cross sections directly depend on the time-dependent compression factor $c=R_0^3/R(t)^3$ or the corresponding
density of the sphere, where $R_0$ is the radius of the uncompressed plutonium core.
The Ansatz
\begin{equation}
n(t,\vec{r}) = f(t) \tilde{n} (t,\vec{r}) \label{sep_ansatz}
\end{equation}
with the inverse bomb period $\alpha(t)$ (Rossi alpha) 
\begin{equation}
f(t) = e^{\Omega (t)} \, , \quad \quad \Omega(t) = \int \limits_{0}^{t} dt \, \alpha(t')dt' \, ,
\end{equation}
leads to
\begin{equation}
\alpha (t) \tilde{n} (t, \vec{r}) =\overline{v} D(t)  \Delta \tilde{n} (t, \vec{r}) + \overline{v}(\nu \Sigma_f (t) - \Sigma_a (t) ) \tilde{n} (t,\vec{r})
-\frac{\partial \tilde{n} (t,\vec{r})}{\partial t}\, . \label{separation}
\end{equation}
In the spherically symmetric case, the Laplacian $\Delta$ becomes
\begin{equation}
\Delta = \frac{\partial^2}{\partial r^2} + \frac{2}{r} \frac{\partial}{\partial r} \, ,
\end{equation}
and making the {\bf{q}}uasi-{\bf{s}}tatic Ansatz where one neglects the last term in Eq. (\ref{separation}) one finds the well-known solution
which is an eigenstate of the Laplacian
\begin{equation}
\tilde{n}_{qs} (t, \vec{r}) \sim  \frac{\sin(B_g(t)r)}{B_g(t) r} \, , \quad r = | \vec{r} |  \le R(t) \, , \label{nf_ansatz}
\end{equation}
where $B_g^2(t)=B_g(t)^2$ is the so-called geometric buckling to be specified below,
and from Eq. (\ref{separation}) and
\begin{equation}
\Delta \tilde{n}_{qs} (t, \vec{r}) = -B_g^2(t)  \tilde{n}_{qs} (t, \vec{r})
\end{equation}
follows the inverse bomb period
\begin{equation}
\alpha (t) = \overline{v} D(t)  (B^2_m (t) - B^2_g (t) )\, , \label{ibp}
\end{equation}
and the so-called material buckling is given in accordance with the literature by
\begin{equation}
B^2_m = \frac{ \nu \Sigma_f - \Sigma_a}{D} = \frac{k_\infty -1}{D} \Sigma_a = \frac{k_\infty -1}{L^2} 
\end{equation}
with $L^2=D / \Sigma_a$. Space and time arguments have been omitted for the sake of notational simplicity.
Above, the infinite multiplication factor for a core without leakage
\begin{equation}
k_\infty = \nu \frac{\sigma_f}{\sigma_a}
\end{equation}
has also been introduced.

Now two comments are in order. First, the separation Ansatz in Eq. (\ref{sep_ansatz}) is certainly not unique.
A separation \`a la $n(t, \vec{r}) = f(t) \tilde{n} (\vec{r})$, such that obviously $\partial \tilde{n} (\vec{r}) / \partial t = 0$,
 is impossible since the diffusion equation under study
is considered on a time-dependent domain $\mathcal{B} (t)=\{ \vec{r} \mid r= | \vec{r} | \le R(t) \}$, and one should note that  $D(t)$ and
$\Sigma_{f,a} (t)$ are time-dependent quantities in $\mathcal{B} (t)$ also whose evolution in time will be governed by the dynamics
induced by the chain reaction to be discussed further below.
Second, we need a strategy to calculate the geometric
buckling appearing in the approximate Ansatz (\ref{nf_ansatz}) for $\tilde{n} (t, \vec{r})$. From diffusion
theory one learns for the time-independent case that the spherically symmetric neutron (flux) density inside a homogeneous ball
with radius $R$ surrounded by a vacuum is maximal in the center ($\vec{r} = \vec{0}$) and minimal at the boundary
($| \vec{r} | = R$) where the neutrons leak out. If $\Phi (\vec{r})=0$ for $| \vec{r} | =R$ were true, the geometric
buckling would be given by $B_g^2 = \pi^2 / R^2$, so that $\tilde{n}_{qs} \sim \sin (\pi R / R)/\pi$ vanishes.
However, according to a common approximation inspired from diffusion theory, the neutron (flux) density vanishes on an imaginary
sphere outside the core with a so-called extrapolated radius $R'=R+\delta$ to be calculated below.
With this boundary condition applied to the time-dependent case, the geometric buckling becomes
\begin{equation}
B_g^2 (t) = \frac{\pi^2}{R' (t) ^2} \, ,
\end{equation}
and $\alpha(t)$ can also be computed.

Integrating over the ball $\mathcal{B} (t)$, we obtain from Eq. (\ref{separation}), disregarding the last term on the right, for the neutron number
\begin{equation}
\frac{dN(t)}{dt} = \alpha(t) N(t) \, , \quad N(t) = \int \limits_{\mathcal{B} (t)} n(t, \vec{r}) \,  d^3 r  \label{growth}
\end{equation}
and consequently
\begin{equation}
N(t)= N_0 e^{\Omega (t)}  \,  , \quad \Omega(t) = \int \limits_{0}^{t} dt \, \alpha(t')dt' \, . \label{diffeq_period}
\end{equation}

The fact that the last term in Eq. (\ref{separation}) $\sim \partial \tilde{n} / \partial t$ is negligible after integration
represents an adiabatic approximation which can be justified by the observation that the neutron dynamics is governed
by a much smaller time scale than the expansion of the plutonium core. The expansion destroys the separability of the
diffusion equation. After having defined how to calculate the quasi-static neutron (flux) distribution and the geometric
buckling in the expanding core as a quasi-static object, Eq. (\ref{diffeq_period}) can be inferred as an
approximation to the true neutron number dynamics.

In the presence of a neutron source (due to spontaneous fission of $^{240}$Pu or an alpha-beryllium
source), Eq. (\ref{growth}) can be equipped with a source term $S(t)$
\begin{equation}
\frac{dN(t)}{dt} = \alpha(t) N(t) + S(t) \, .
\end{equation}

Using now the extrapolated radius $R'$ inspired from neutron transport theory \cite{WeinbergWigner} where the
extrapolated neutron flux density Eq. (\ref{nf_ansatz}) is assumed to vanish
\begin{equation}
R' = R + \delta = R + 0.71045 \lambda_{tr} \label{extrapol}
\end{equation}
leads in the truly static case ($\alpha=0$) to the condition for the critical extrapolated radius $R'_c$
which is related to the geometric buckling introduced above
\begin{equation}
B_g^2 = \frac{\pi^2}{R_c^{'2}} = B_m^2 \, .
\end{equation}

For illustrative purposes, we readily calculate the critical mass $m^{49}_c$ for pure plutonium in the alpha phase.
The neutron cross sections averaged over the fission spectrum are given by
$\sigma_f=1.8\rm b$, $\sigma_s=\sigma_{nn}+\sigma_{nn'}= 4.566 \rm b + 1.369 \rm b= 5.935 \rm b$,
$\sigma_c=\sigma_{n \gamma}=0.065 \rm b$, $\sigma_{t}=7.8 \rm b$, and $\sigma_a=\sigma_t - \sigma_s=1.865 \rm b$.
The average number of fast neutrons released in a fission induced by fast fission neutrons is $\nu=3.091$,
delayed neutrons included in the stationary case.
In the alpha phase, the density of plutonium is $\rho_\alpha^{49}= 19.86 \cdot 10^3 \rm kg \cdot m^{-3}$,
from the isotope mass $M^{49}=239.052 \rm u$ one calculates the density of plutonium atoms
$\rho_A^{49}=5.003 \cdot 10^{28} \rm m^{-3}$ with the Avogadro constant $N_A=6.02214 \cdot 10^{23}$.
From diffusion theory follow the approximate expressions for
the diffusion constant $D$ and the  transport mean free path $\lambda_{tr}$
\begin{equation}
D=\frac{1}{3 ( \Sigma_t - \overline{\mu}_0 \Sigma_s)} = \frac{\lambda_{tr}}{3} = 0.856 \rm cm
\end{equation}
with an approximate value of the average cosine of the neutron scattering angle $\overline{\mu}_0 = 2/(3A)$,
where $A=239.05$ is the atomic mass number.
One finally obtains
\begin{equation}
R_c=R'_c - \delta = 6.757 {\rm cm} - 1.824 {\rm cm} = 4.933 {\rm cm} \, ,
\end{equation}
corresponding to a critical mass of
\begin{equation}
m_{c, \alpha}^{49} = \frac{4 \pi}{3} \rho^{49} R_c^3 = 10.0 \rm kg \, .
\end{equation}
An analogous calculation for pure $^{23 {\bf{5}}}_{\, \, 9 {\bf{2}}}$U with
$\sigma_f=1.219\rm b$, $\sigma_s=\sigma_{nn}+\sigma_{nn'}= 4.409 \rm b + 1.917 \rm b= 6.326 \rm b$,
$\sigma_{n \gamma}=0.095 \rm b$, $\sigma_{t}=7.64 \rm b$, $\sigma_a=\sigma_t - \sigma_s=1.314 \rm b$,
$\rho^{25}=18.9 \cdot 10^3 \rm kg \cdot m^{-3}$, $\rho_A^{25}=235.044$, and $\nu=2.583$ gives 
$R_c = 8.092\rm cm$ and $m_c^{25}=41.9 \rm kg$. Analogous calculations concerning the critical mass of
uranium can also be found in \cite{Reed}.

Interestingly, doubling the density of $^{239}$Pu by compressing it by a compression factor $c=2$ reduces the critical radius to
$2.47 \rm cm$ and the critical mass to $2.50 \rm kg$.
Adding an extended neutron reflecting tamper around the plutonium core also lowers the critical mass.
In this case, the spherically symmetric neutron flux density in the tamper behaves like $e^{-\kappa r}/r$,
and the critical mass can be calculated from the critical condition that the stationary diffusion current density is continuous everywhere.

\subsection{Kinetics}
We want to generalize the preliminary considerations above to a homogeneous $^{239}$Pu sphere with a time-dependent
radius $R(t)$ containing $^{23{\bf{9}}}_{\, \, 9{\bf{4}}}$Pu with a total mass $m^{49}(t)$.
It is straightforward to calculate relevant quantities like macroscopic cross sections
discussed above from convenient quantities like
the number of $^{239}$Pu atoms $N^{49}(t)$, the volume $V(t)$ of the sphere or the corresponding atomic density $\rho_A^{49} (t)$ 
\begin{equation}
\rho_A^{49} (t) = \frac{N^{49} (t)}{V(t)} \, , \quad N^{49} (t)= \frac{m^{49} (t)}{M} N_A \, , \quad V(t)=\frac{4 \pi}{3} R(t)^3 \, ,
\end{equation}
where $V(t)$ is the momentary volume of the sphere. Before the chain reaction starts, the mass and the number of $^{239}$Pu
nuclei is given by initial values $m_0^{49}$ and $N_0^{49}$, respectively.
From $R(t)$ and $\lambda_{tr} (t)$ follows the extrapolated radius
$R'(t)$ and the corresponding material and geometric bucklings $B_{m,g}^2 (t)$; in the case of the geometric buckling,
the quasi-static approximation is
\begin{equation}
B_g^2 (t) = \frac{\pi^2}{R'(t)^2}
\end{equation}
such that $\tilde{\Phi}  (t, R') = 0$.
The inverse bomb period $\alpha (t)$ follows from Eq. (\ref{ibp}) and serves for the update of the neutron flux density
inside the sphere. With respect to the inverse bomb period, it is useful to introduce the (time-dependent) prompt neutron generation time
\begin{equation}
\Lambda = (\nu \Sigma_f \overline{v})^{-1} \, ,
\end{equation}
the effective neutron multiplication factor $k$ and the reactivity $\rho$
\begin{equation}
\rho = \frac{k-1}{k} \, ,
\end{equation}
which are related to the inverse bomb period via
\begin{equation}
\alpha = \rho / \Lambda =  (k -1) / \tau \, ,
\end{equation}
where $\tau = k \Lambda$ is
the average neutron lifetime in the plutonium assembly.
Then the $N(t)$ neutrons inside the sphere generate an average flux density of
\begin{equation}
\overline{\Phi}(t) = \overline{n}(t)  \overline{v} = \frac{N(t)}{V(t)} \overline{v} \, ,
\end{equation}
and the power released inside the sphere is
\begin{equation}
P(t)=\Sigma_f(t) N(t) \overline{v} \epsilon_f \, ,
\end{equation}
where $\epsilon_f = 200 \rm MeV$ is the average energy released per fast fission (anti-neutrinos neglected).
The expansion of the sphere is driven by radiation pressure in the 'hot phase' of the explosion.
Assuming that the released energy
\begin{equation}
E(t)= \int \limits_0^t dt' \, P(t')
\end{equation}
is converted into black body radiation and kinetic energy of the expanding matter shell,
one has an average pressure inside the sphere
\begin{equation}
p(t) = \frac{E_c (t)}{3 V(t)} = \frac{4 \sigma}{3 c} T(t)^4 \, ,
\end{equation}
with the Stefan-Boltzmann constant $\sigma = 5.67 \cdot 10^{-8} \rm W \cdot m^{-2} \cdot K^{-4}$ and the
speed of light $c=299792458 \rm m \cdot s^{-1}$.
$E_c(t)$ is the photon energy stored inside the sphere. 
Also the decrease of the number $N^{49}$ of $^{239}$Pu atoms should be taken into account
\begin{equation}
N^{49} (t) =N_0^{49} - E(t)/\epsilon_f \, .
\end{equation}
Since the pressure gradient drives the expansion of the sphere, one has to make a reasonable assumption at the present
stage of the model concerning the behaviour of the pressure gradient.
However, since the pressure will create a fireball filled with a photon gas pushing away all the
matter from the explosion center, the corresponding radiation energy is converted into kinetic energy of the matter,
and an Ansatz describing the work done by the radiation pressure on the surrounding and moving matter distributed over an
area of $4 \pi R(t)^2$ according to
\begin{equation}
\frac{d E_{kin}(t)}{dt} = \frac{1}{2} \frac{d}{dt} (m(t) \dot{R}(t)^2) = 4 \pi R(t)^2 p(t) \dot{R} (t)
\end{equation}
is reasonable, where $m(R(t))$ is the mass of all the matter that was located inside a sphere with radius $R(t)$ before the
nuclear detonation and that got compressed onto a spherical shell with approximately the same radius.
Accordingly, the photon energy inside the sphere is given by
\begin{equation}
E_c (t) = E(t) - E_{kin} (t) \, .
\end{equation}

\FloatBarrier
\section{Simulation results}
\subsection{Trinity implosion bomb}
Since in the Trinity test,
the plutonium core was (probably) surrounded by a $108 \rm kg$ $^{235}$U-tamper, a $130 \rm kg$ aluminum
pusher shell plus $4430 \rm kg$ additional material including the high explosives surrounding the shells around the core,
$m(R)$ was modeled by
\begin{displaymath}
m(R)= 6.2 {\rm kg} + 108 {\rm kg} \frac{R(t)^3-R_{min}^3}{R_T^3-R_{min}^3} \Theta(R_T-R(t))
\end{displaymath}
\begin{displaymath}
+\biggl( 108 {\rm kg} + 130 {\rm kg}  \frac{R(t)^3-R_T^3}{R_{Al}^3-R_T^3} \biggr) \Theta(R_{Al}-R(t)) \Theta(R(t)-R_T)
\end{displaymath}
\begin{displaymath}
+\biggl( 238  {\rm kg} + 4432 {\rm kg} \frac{R(t)^3-R_{Al}^3}{R_B^3-R_{Al}^3} \biggr) \Theta(R_B-R(t)) \Theta(R(t)-R_{Al})
\end{displaymath}
\begin{equation}
+\bigl (4670 {\rm kg} + \frac{4 \pi}{3}  (R(t)^3-R_B^3)  \rho_{Air} \bigr) \Theta(R(t)-R_B)
\end{equation}
with the minimum radius $R_{min}$ of the imploded core, a tamper radius $R_T=11.1 {\rm cm}/c^{1/3}$,
an aluminum pusher shell radius $R_{Al}=23.5 {\rm cm} / c^{1/3}$, an approximate bomb radius when the nuclear
detonation starts $R_{B}=1 {\rm m}$, and the density of air $\rho_{Air}=1.29 {\rm kg \cdot m^{-3}}$. Above, approximate
(de)\-com\-pression effects of the bomb material when reaching a
maximum core compression factor $c$ have been taken into account.

\begin{figure}[!htbp]
\begin{center}
\includegraphics[width=11.6cm]{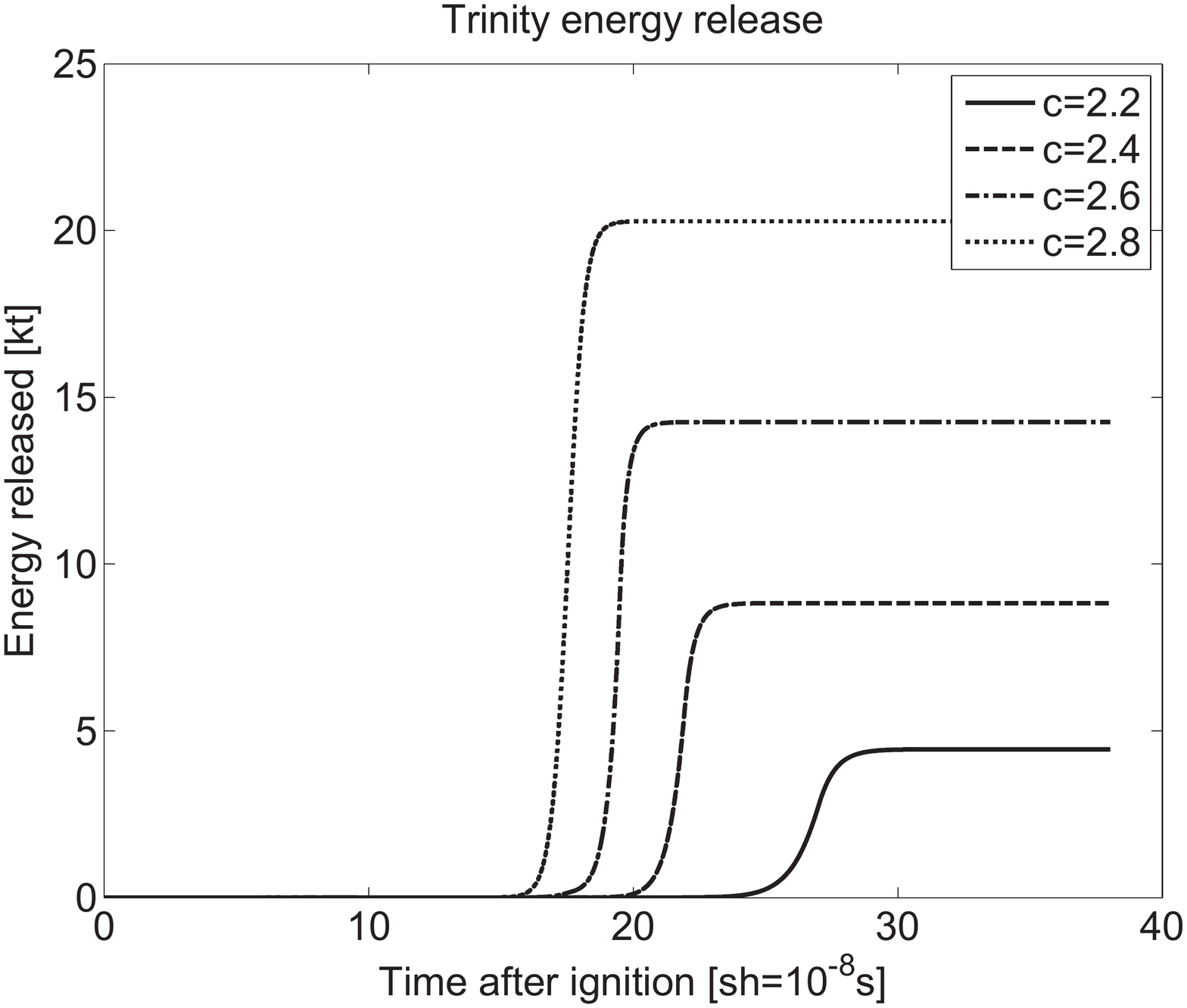}
\caption{Explosion yield for the Trinity core for different maximum compression factors.}
\label{fig_compression_yield}
\end{center}
\end{figure}

The small neutron source located in the center of the plutonium core
starting the chain reaction at maximum compression was assumed to emit enough neutrons of the order $10^8 \rm s^{-1}$
such that the chain reaction is initiated immediately when the shock wave of the conventional explosion reaches
the center of the core.
Since delayed neutrons play no role on the time-scale of a nuclear explosion, an average number of $\nu=(1-0.0021) \cdot 3.091$
prompt neutrons per fission was used for the simulation. Using super-grade plutonium, the probability that the chain reaction
starts due to spontaneous fission of $^{240}$Pu is rather small for core compression times of the order of $10^{-5} \rm s$.

The Trinity plutonium core consisted of $6.2 \rm kg$ plutonium-gallium alloy containing $0.9 \rm wt. \%$ gallium
with a fast neutron scattering cross section of $3.0 \rm b$. The plutonium phase with the lowest density is the delta phase, with theoretical
density of $15.92 {\rm g \cdot cm^{-3}}$. Although existing only in the temperature range of $310 - 452^o {\rm C}$, it
can be stabilized at room temperature by adding small quantities of gallium to the plutonium.

The convenient non-metric units of a kiloton ($1 {\rm kt} = 10^{12} {\rm cal} = 1.31 \cdot 10^{23}$ fission
events) and a shake ($1 {\rm sh} = 10^{-8} \rm s$) will also be used in the following \cite{Sandmeier}.

Most of the results presented in this section were calculated for a chain reaction starting at the time of maximum core compression
with a compression factor of $2.5$ \cite{Semkow}. Adaptive timesteps of the order of $10^{-12} \rm s$ have been
used to model the actual explosion based on a simple Euler discretization of the dynamical
equations presented in the last section.

One should note that it does not make too much sense to calculate an extrapolated radius for the neutron
flux density outside the plutonium core for the uranium tamper during the intense phase of the nuclear explosion,
since the average lifetime $\tau^{28}$ of the neutrons in the $^{23{\bf{8}}}_{\, \, 9{\bf{2}}}$U tamper is much bigger than in the core.
The lifetime $\tau^{28}$ is given by the average length of the path $\lambda_{28}$ a neutron travels in the uranium until it gets absorbed
there by fission or capture
\begin{equation}
\lambda^{28} = \frac{1}{\Sigma_a^{28}} = \frac{1}{\Sigma_f^{28} + \Sigma_c^{28}}= \frac{1}{\rho_A^{28} ( \sigma_f^{28} + \sigma_c^{28})}
\end{equation}
divided by the neutron velocity. The corresponding cross sections averaged over the fission spectrum are
$\sigma_f^{28} \simeq 0.3 \rm b$ and $\sigma_c^{28} \simeq 0.07 \rm b$, and from the uranium density $\rho^{28} = 18.95 \rm g \cdot cm^{-3}$
at room temperature and pressure and a compression factor $c$
one calculates $\tau^{28} \simeq 3.13 \cdot 10^{-8} {\rm s}/c$.
Therefore, the neutrons need a time to adapt their distribution which is
of the order of the time the main part of the energy is released by the nuclear explosion. In fact, it turns out that
simply using the extrapolated length for a core surrounded by a vacuum calculated from Eq. (\ref{extrapol}) already 
leads to reasonable explosion yields for Trinity, as depicted in Fig. (\ref{fig_compression_yield}).
For a compression factor of $c=2.63$, the yield is $15.1$ kilotons.

\begin{figure}[!htbp]
\begin{center}
\includegraphics[width=11.0cm]{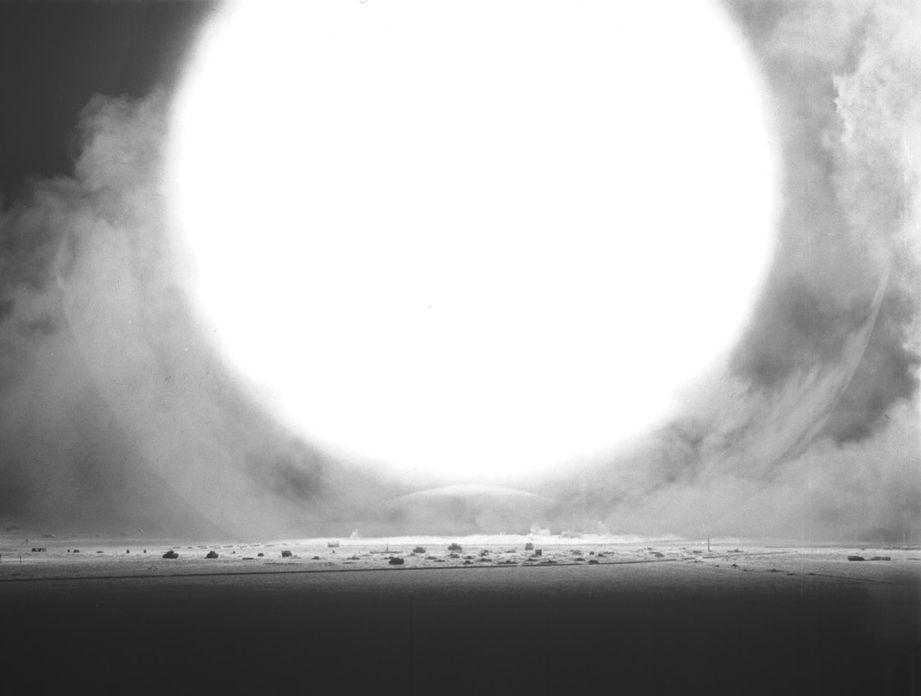}
\caption{Image of the Uphot-Knothole Grable fireball. Note the reflection of the shock wave
near the ground.}
\label{fig_upshot_grable}
\end{center}
\end{figure}

\begin{figure}[!htbp]
\begin{center}
\includegraphics[width=11.0cm]{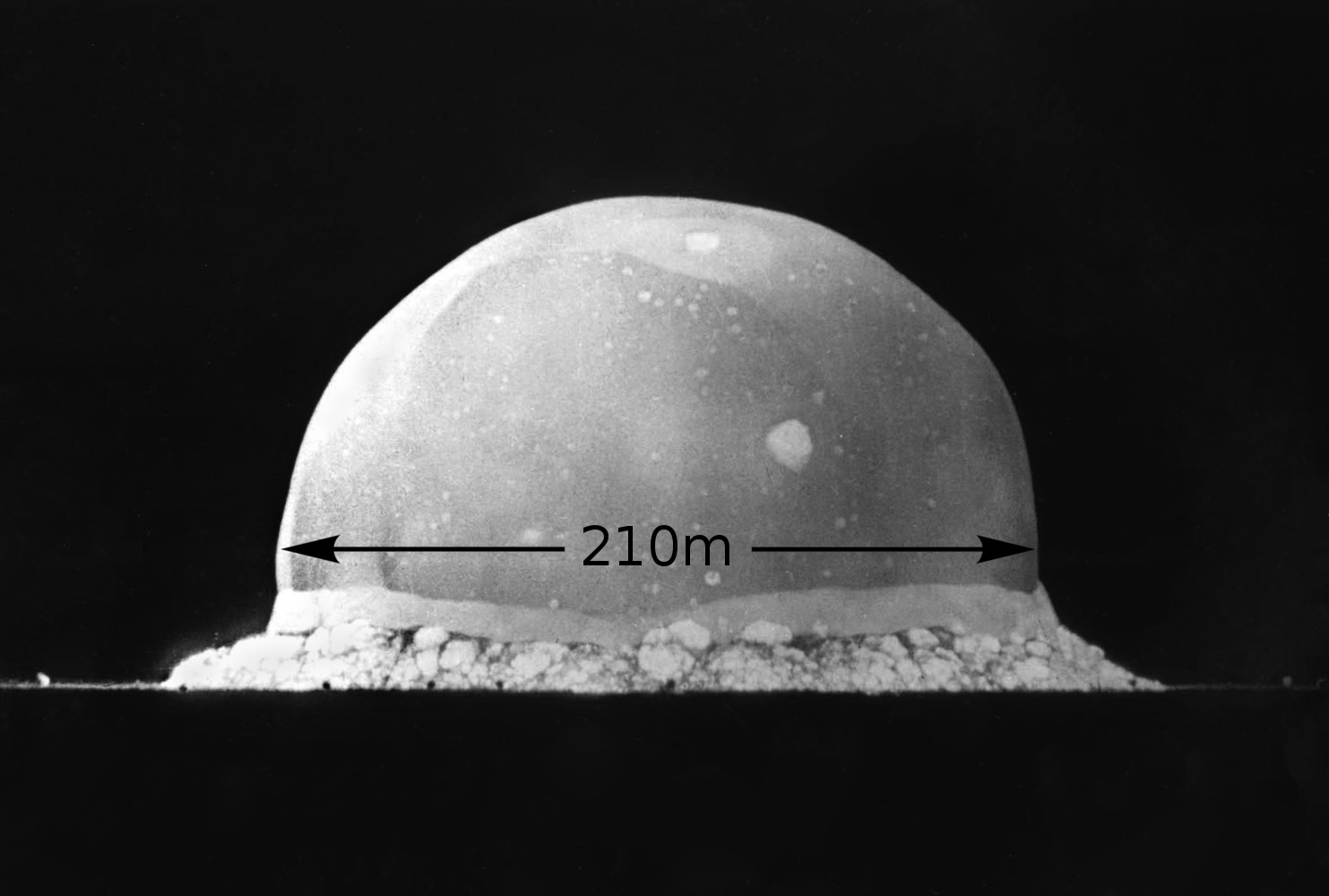}
\caption{Trinity fireball 0.016s after ignition.}
\label{fig_trinity}
\end{center}
\end{figure}

\begin{figure}[!htbp]
\begin{center}
\includegraphics[width=11.6cm]{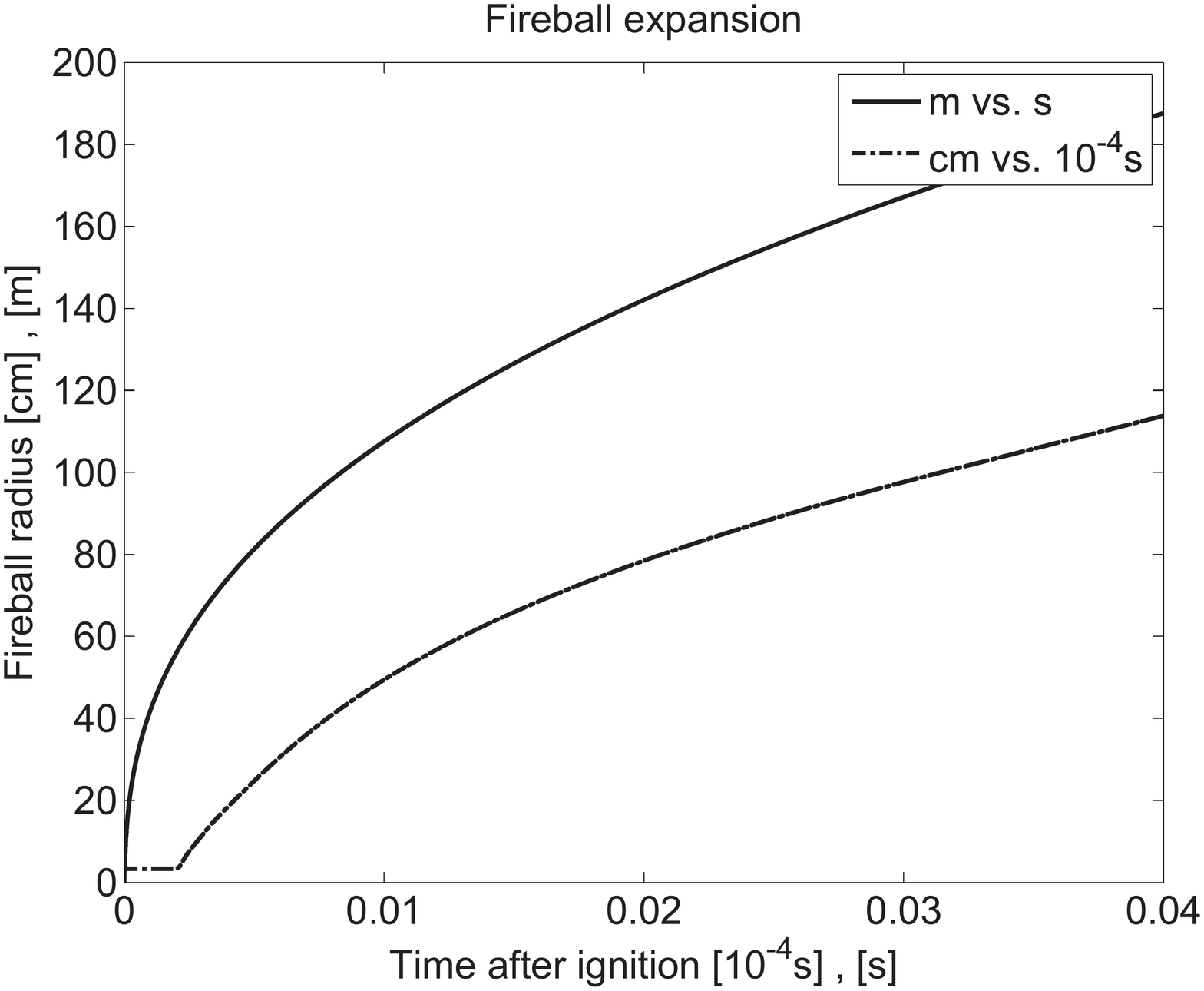}
\caption{Expansion of the fireball shortly after ignition.}
\label{fig_trinity_expansion}
\end{center}
\end{figure}

However, in order to have consistency with some data used in the literature, the constant $0.71045$
in Eq. (\ref{extrapol}) was slightly reduced to $0.633$ in order to gauge our model to a yield of $15 \rm kt$ for
$c=2.5$. Since the limits of the physical validity of diffusion theory are reached and since the point kinetic model is based
on some strong homogeneity assumptions concerning the neutron distribution inside the core in the prompt supercritical
phase and the idea of a spherical shell-like structure of the fireball, such a strategy may be acceptable.
A yield of 15kt is the currently accepted value for the Trinity test due to the $^{239}$Pu content in the Gadget
only; still, it is estimated that the $^{238}$U tamper additionally released about $6 \rm kt$, but we do not try to model
this surplus in this study which finally leads to the generally accepted value for the total Trinity yield of of $21 \rm kt$.
Of course, the less relevant energy production by the tamper could also be assessed in our approach by modeling
the prompt neutron leakage from the core into the tamper. In the {\emph{static}} and homogeneous case,
the neutron leakage rate $\mathcal{L}$ is given by
\begin{equation}
\mathcal{L}= D B^2 \overline{v} N
\end{equation}
with the geometric buckling
\begin{equation}
B^2 = - \frac{\Delta \Phi}{\Phi} \,  .
\end{equation}
\begin{figure}[!htbp]
\begin{center}
\includegraphics[width=11.6cm]{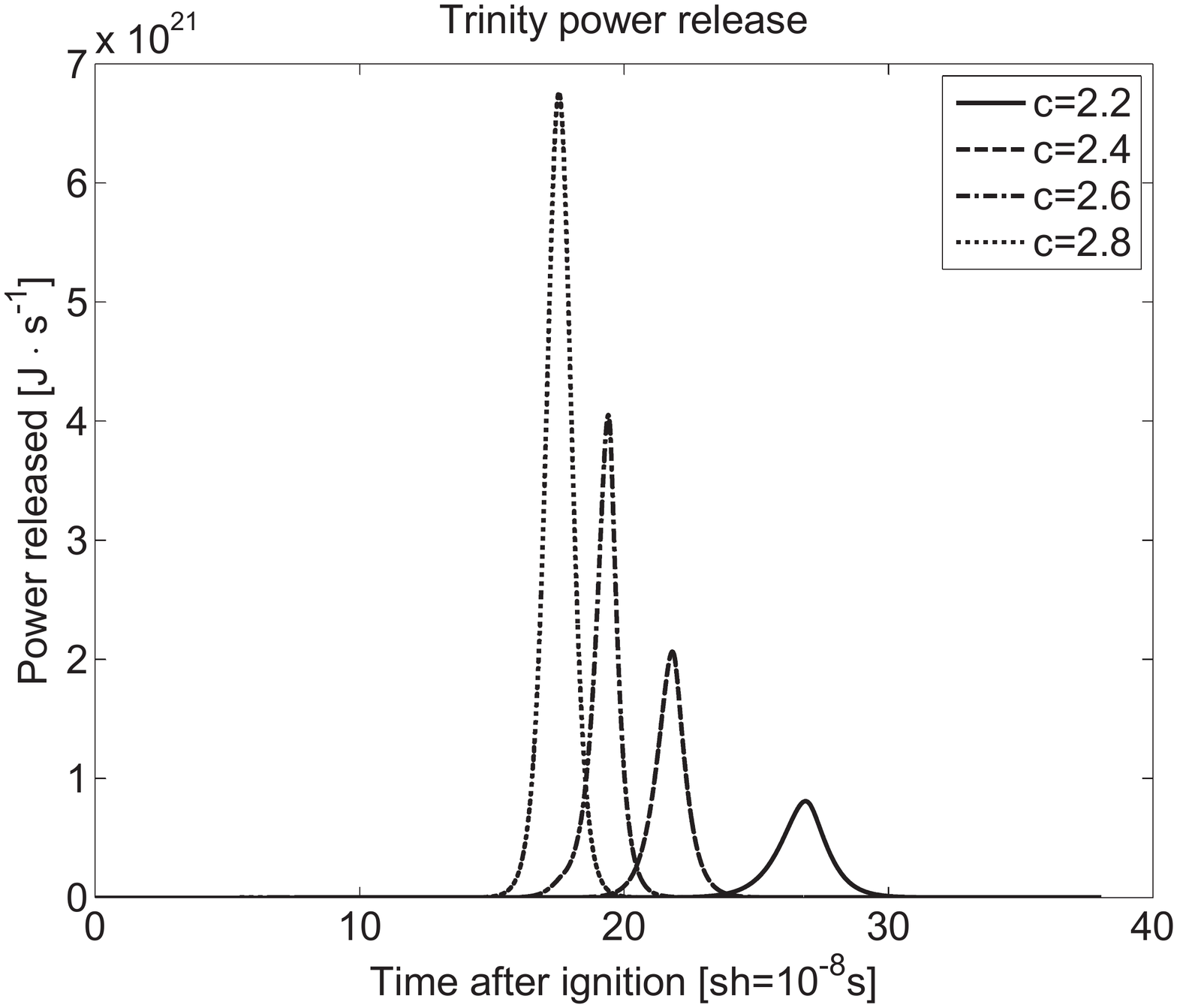}
\caption{Total power released by the Trinity core for different maximum compression factors.}
\label{fig_power_yield}
\end{center}
\end{figure}
Fig. (\ref{fig_trinity_expansion}) displays the size of the fireball as a function of the time after
ignition. The results are in very good agreement with the actual radii measured in the Trinity test -
see, e.g., Fig. (\ref{fig_trinity}) showing a high-speed rapatronic camera photograph of the fireball
taken $0.016 \rm s$ after ignition \cite{nwa}.
One should note at this stage that the total yield $Y$ of the nuclear explosion can be decomposed in a more
elaborate model into different parts. E.g., within a millionth of a second after the explosion, all matter including
the bomb itself and the surrounding air is transformed into a very hot plasma which emits thermal radiation as
X-rays, which again gets reabsorbed by the dense shock front of the fireball itself. After some seconds, the energy
of the explosion can be decomposed into the blast energy $Y_B$, i.e. the kinetic energy transferred primarily to the air
(still about $50 \%$), thermal radiation $Y_{TR}$ including light ($\simeq 35 \%$), and nuclear radiation $Y_{NR}$
of various types ($\simeq 15 \%$). At this time, the focus of investigations in the literature is rather on the
damage done to humans and the environment
and the present considerations must be replaced by different physical and ethical concepts \cite{Glasstone}, \cite{Jungk}.
Fig. (\ref{fig_upshot_grable}) shows a snapshot taken in 1953 shortly after ignition of the device used in the nuclear weapons test 
Upshot-Knothole Grable, where an estimated 15kt gun-type fission bomb exploded $160 \rm m$
above ground, producing finally a spherically symmetric fireball, despite the asymmetric aspects of the pre-ignition device
\cite{nwa}.
Supposing that a considerable part $Y_B$ of the total yield $Y$  of an atomic bomb is converted into the
kinetic energy of the matter moving at a distance $R(t)$ from the explosion center with a velocity
$\dot{R}(t)$, and neglecting the mass of the bomb for fireballs with a radius larger than $\simeq 20 \rm m$
and the energy of the radiation inside the ball, one has from
\begin{equation}
m(R) \simeq \frac{4 \pi}{3} R^3 \rho_{Air}
\end{equation}
the kinetic energy which equals the blast energy
\begin{equation}
Y_B = \frac{1}{2} m(R(t)) \dot{R} (t)^2 = \frac{2 \pi \rho_{Air}}{3} R(t)^3 \dot{R} (t)^2 \, ,
\end{equation}
and therefore
\begin{equation}
\dot{R} (t) R(t)^{3/2} = \frac{2}{5} \frac{d}{dt} R(t)^{5/2} = \sqrt{\frac{3 Y_B}{2 \pi \rho_{Air}}} \, .
\end{equation}
Integrating this expression leads to
\begin{equation}
R(t)^{5/2} = \sqrt{\frac{75 Y_B}{8 \pi \rho_{Air}}} (t - t_0)
\end{equation}
with some integration constant $t_0$, and finally one has
\begin{equation}
R(t) =  \biggl( \frac{75 Y_B}{8 \pi \rho_{Air}} \biggr)^{1/5} (t - t_0)^{2/5} \, . \label{blast_radius}
\end{equation}
For large radii, the integration constant is negligible and Eq. (\ref{blast_radius}) is in good agreement
with observations from nuclear tests and in very good agreement with the numerical results presented
in Fig. (\ref{fig_trinity_expansion}). One also observes that doubling the blast yield by a factor of two
increases the radius $R \sim Y_B^{1/5}$ of the fireball at the same time after ignition only by $(2^{0.2}-1)=15 \%$.

\begin{figure}[!htbp]
\begin{center}
\includegraphics[width=11.6cm]{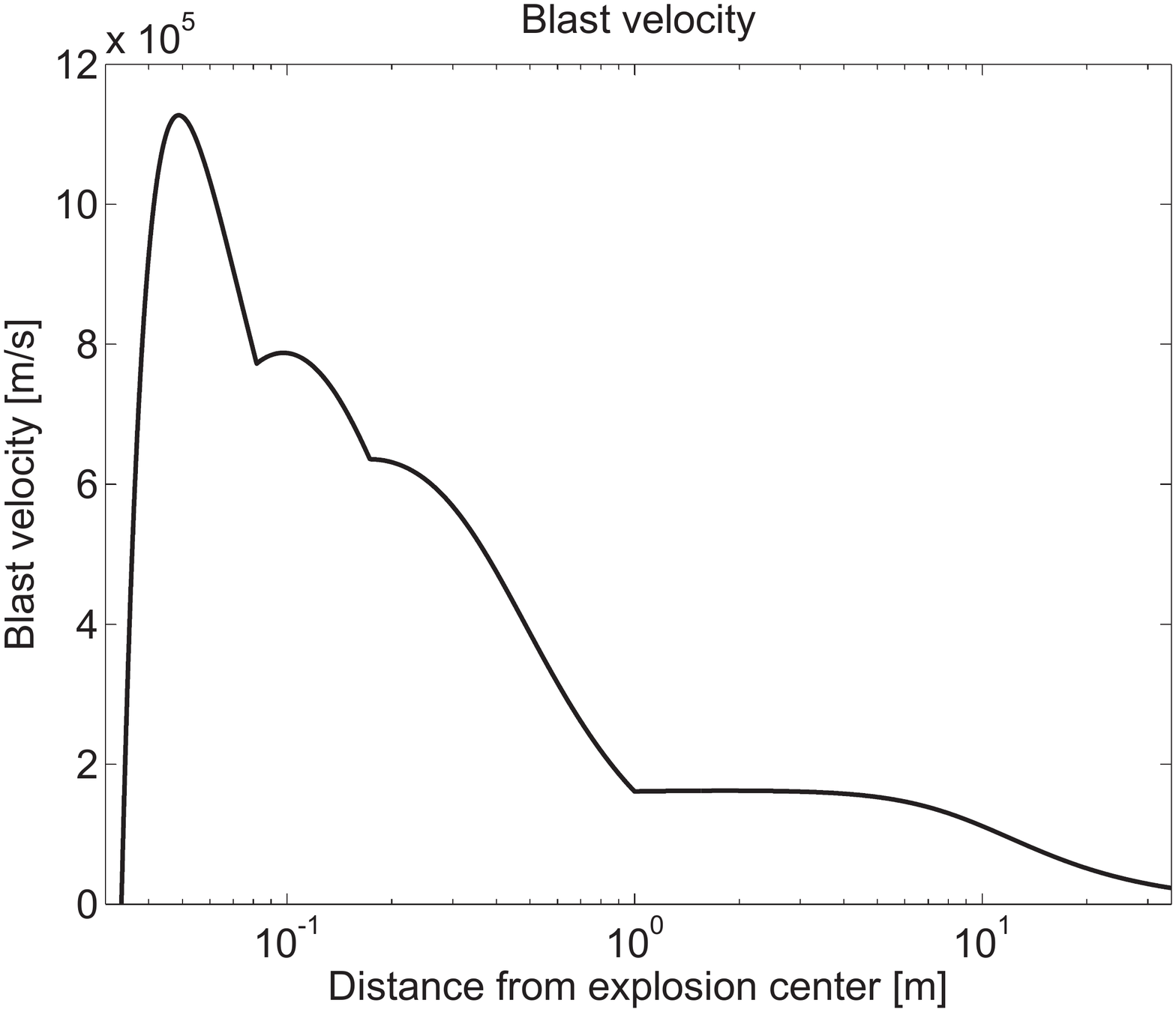}
\caption{Expansion of the matter shell shortly after ignition.}
\label{fig_exp_velocity}
\end{center}
\end{figure}

\begin{figure}[!htbp]
\begin{center}
\includegraphics[width=11.6cm]{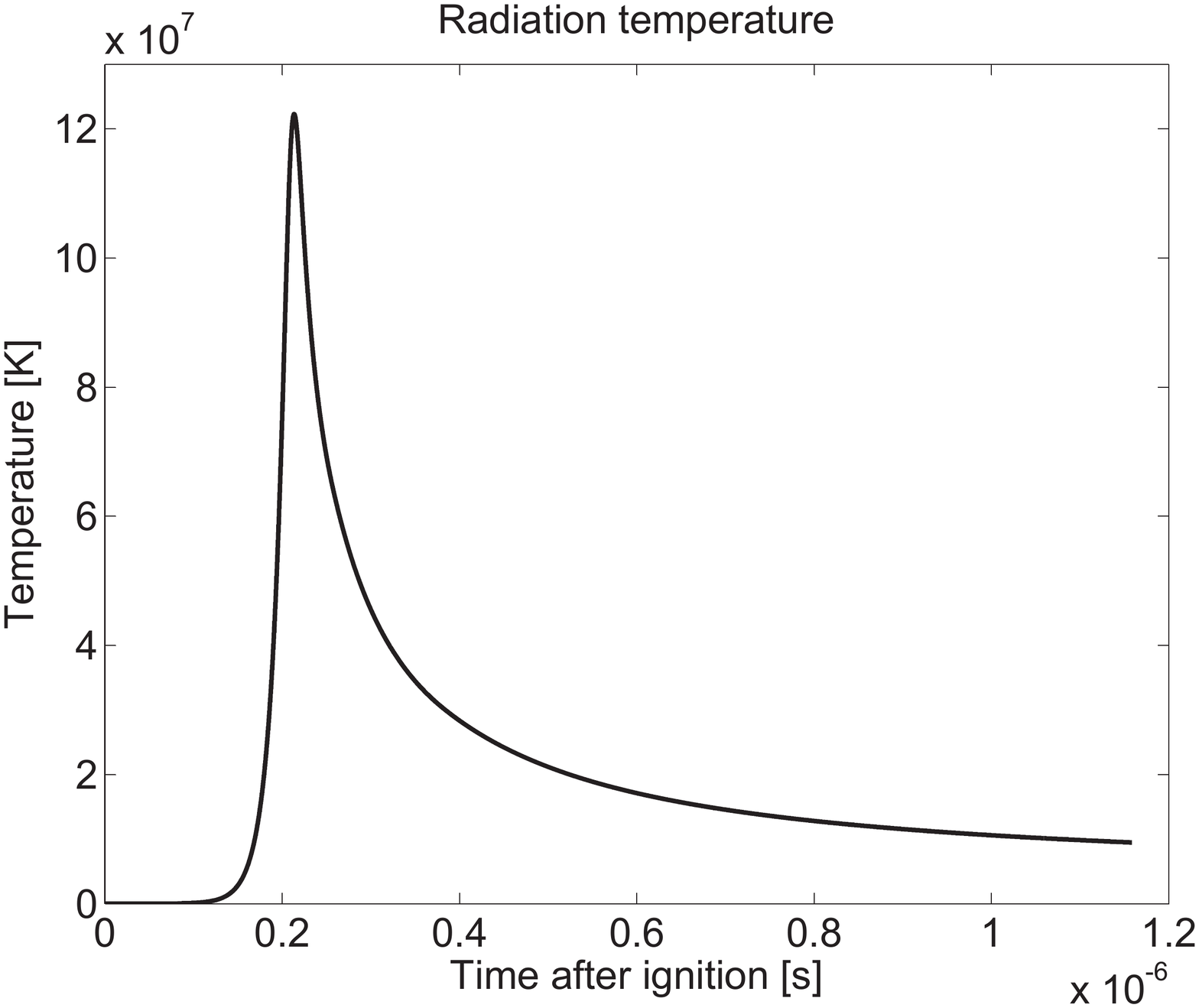}
\caption{Temperature of the black body radiation inside the fireball.}
\label{fig_trinity_temperature}
\end{center}
\end{figure}

It is also instructive to consider the fireball expansion velocity as a function of the distance from the
explosion center. The tail for larger distances in Fig. (\ref{fig_exp_velocity}) corresponds to the description
via Eq. (\ref{blast_radius}). After a first acceleration phase due to the violent energy release by the plutonium
core, the expansion enters a deceleration phase caused by the high density of the uranium tamper.
When the explosion enters the aluminum with a lower density, the pressure is still  high enough to initiate
a reacceleration of the expansion. Finally, the rest of the bomb and the air surrounding the device are
pushed away, still at a speed of the order of $100 \rm km \cdot s^{-1}$.
Fig. (\ref{fig_trinity_temperature}) displays the radiation temperature reached inside the Trinity fireball.
Temperatures of the order of $10^8 \rm K$ correspond to particle energies of
$k_B \cdot 10^8 {\rm{K}} =  8.62 \rm keV$ with Boltzmann's constant $k_B=1.38 \rm J \cdot K^{-1}$.
The temperature for complete ionization of plutonium is of the order of some $10^{9} \rm K$ and the binding
energy of the innermost K-shell electrons is larger than $10^5 \rm eV$, therefore,
in the plasma of a fission explosion one is still far away from complete ionization, but the plutonium
atom may lose a large fraction of its electrons for a very short period of some few shakes. Note that
for partially ionized plasma the degree of ionization can be calculated from the Saha equation \cite{Saha}.
An analysis of the numerical results of the point kinetic model confirmed that the total yield $Y$ of the explosion
of a plutonium implosion device is approximately proportional to the inserted reactivity $\rho_{max}$, i.e.,
the reactivity of the compressed core when the chain reaction starts, as anticipated by Serber \cite{Serber}.
Roughly, one has in the present model
\begin{equation}
Y = 128 {\rm kt \cdot kg^{-1}} m_0^{49} \rho_{max}^3 \, .
\end{equation}
Of course, the total yield depends on the physical parameters governing the point kinetic model.
The dependence of the yield is not very pronounced for variations of $\epsilon_f$. If more energy would be released per
single fission, the core would expand faster and reach second prompt criticality within a shorter period, which has
a negative impact on the total yield of the device. However, it is advantageous to have fast neutrons,
large fission cross sections, and a large fast fission neutron yield $\nu$.

The fraction of plutonium nuclei undergoing fission during a nuclear explosion is called the efficiency of the bomb.
Only about $13 \%$ of the plutonium in the core was destroyed in the Trinity test.
In the vast majority of cases, a plutonium nucleus splits into two smaller nuclei and 2,3, or 4 additional neutrons.
Only about $0.2 \% - 0.4 \%$ of fissions are ternary fissions, producing a third light nucleus such as $^4$He ($90 \%$)
or $^3$H ($7 \%$). E.g., fissioning 1000 $^{239}$Pu nuclei produces on an average $72$ fission fragments with an atomic mass number
of $135$ - namely $^{135}_{\, \, 51}$Sb, $^{135}_{\, \, 52}$Te, and $^{135}_{\, \, 53}$I. These fragments are neutron-rich
and tend to undergo subsequent $\beta^-$-decays according to the decay chain with given half-lives
\begin{equation}
^{135}_{\, \, 51} {\rm Sb} \,  \underset{1.7 {\rm s}}{\longrightarrow} \, ^{135}_{\, \, 52} {\rm Te}
 \,  \underset{19 {\rm s}}{\longrightarrow} \,  ^{135}_{\, \, 53} {\rm I}  \,  \underset{6.6 {\rm h}}{\longrightarrow} \, 
^{135}_{\, \, 54} {\rm Xe }  \,  \underset{9.1 {\rm h}}{\longrightarrow} \,  ^{135}_{\, \, 55} {\rm Cs} 
\end{equation}
until a (quasi) stable fission product like $^{135}_{\, \, 55}$Cs with a half-life of $2.3 \cdot 10^6$ years is reached.
During a nuclear explosion, such $\beta^-$-decays of fission fragments  play no role due to the very short time-scale
of the nuclear chain reaction. Still, the initial fission fragments have some influence on the chain reaction.
Because they are neutron-rich, their absorptive properties are rather irrelevant, but the fragments act as additional scatterers to the neutrons,
influencing thereby the neutron transport inside the core. Scattering has a confining effect, since the neutron diffusion constant
decreases when the neutron scattering cross section increases. The scattering cross sections of the fission fragments were not
taken into account in the present simulation, but including a corresponding macroscopic scattering cross section term in the
point kinetic model is straightforward.

Note, however, that Xenon-135 has the highest known thermal neutron absorption cross section of any nuclide, namely
$\sigma_a = 2.65 \cdot 10^6 \rm b$ for neutrons with a kinetic energy of $0.025 \rm eV$.
In nuclear reactors, $^{135}$Xe can strongly influence the reactivity balance, but its concentration typically varies
on a time-scale of the order of some hours.

\subsection{Fizzles}
Reactor grade plutonium contains different plutonium isotopes.
Whereas the heat generated by the $^{238}$Pu is a problem for the integrity of the explosives
surrounding the nuclear part of the bomb, $^{240}$Pu is a source of fast neutrons which can start
the chain reaction before the plutonium core has reached sufficiently high compression for the intended efficiency.

$^{240}$Pu undergoes $479.1 \pm 5.3$ spontaneous fissions per gramsecond,
releasing neutrons at a rate of about $(1032 \pm 14) {\rm g^{-1} \cdot s^{-1}}$ \cite{Goettsche}.
In most cases, the splitting nucleus emits one, two, or three neutrons with comparable probabilities.
When a neutron is released in the core, there is still some probability that it escapes the fissile zone.
The probability that a fast neutron triggers a fission is given by $P_f = k_{eff}/\nu$. In the limit of an infinitely extended reactor,
this probability becomes $P_{f,\infty}=k_\infty/\nu=\sigma_f/\sigma_a$. In a critical $^{239}$Pu assembly where $k=1$, one has
$P_{f,crit}=1/\nu$, i.e. only one third of the neutrons initiates a fission which releases about three new neutrons
on an average. A small part of the neutrons will be captured, producing thereby $^{240}$Pu, but most of the neutrons
diffuse out of the core. This fact has to be taken into account when modeling fizzle probabilities.

For the sake of clarity, we mention here that fissile materials can sustain a chain reaction with neutrons of any energy, whereas
fissionable materials are materials that can only be made to fission with fast neutrons.
Fertile materials are materials that can be transformed (i.e., transmuted) into fissile materials by the bombardment of
neutrons inside a reactor. In this sense, $^{240}$Pu is fissionable and fertile.

\begin{figure}[!htbp]
\begin{center}
\includegraphics[width=11.0cm]{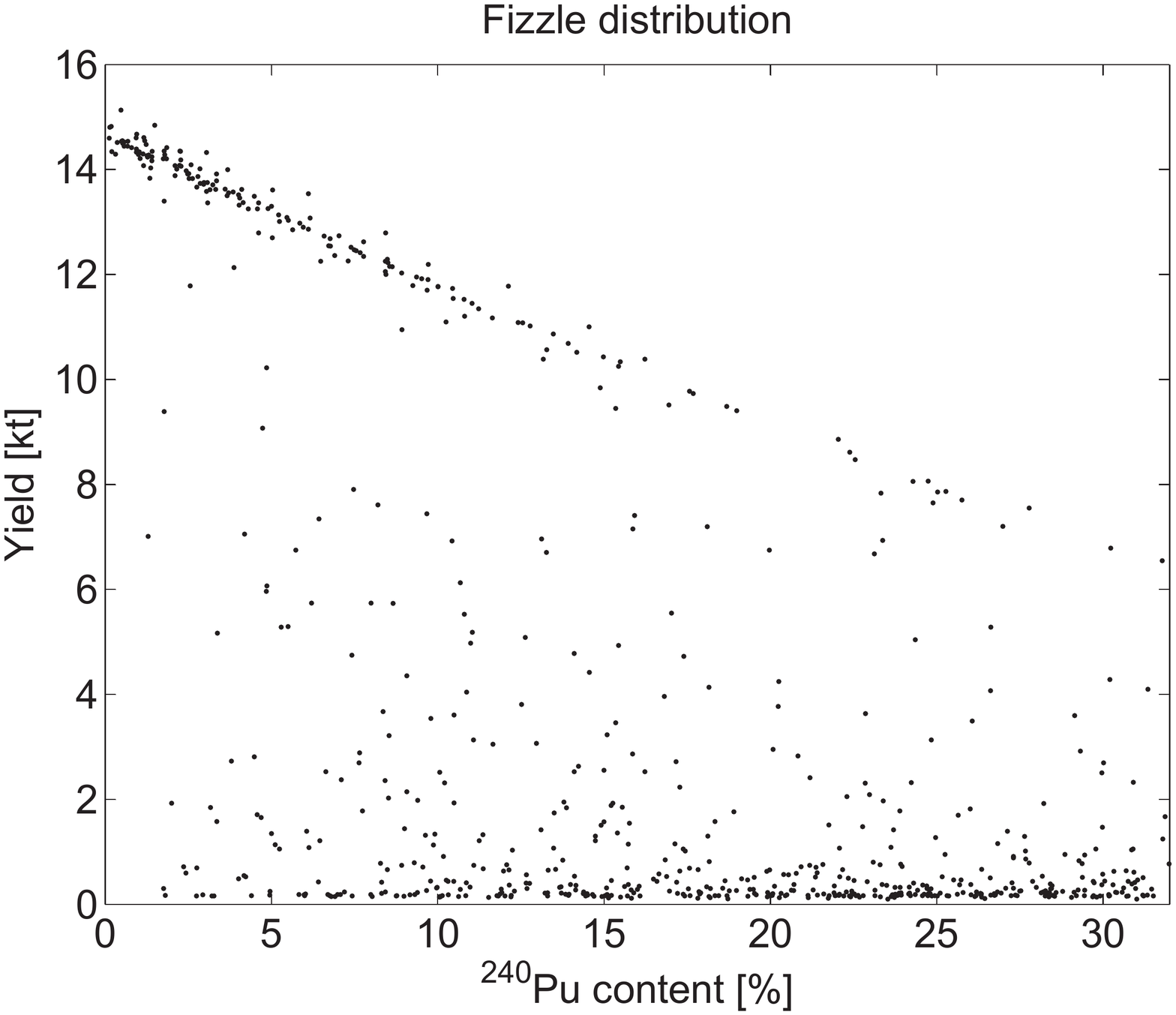}
\caption{Yield distribution for a randomly chosen $^{240}$Pu content (Trinity type device).}
\label{fig_trinity_fizzle}
\end{center}
\end{figure}

\begin{figure}[!htbp]
\begin{center}
\includegraphics[width=11.0cm]{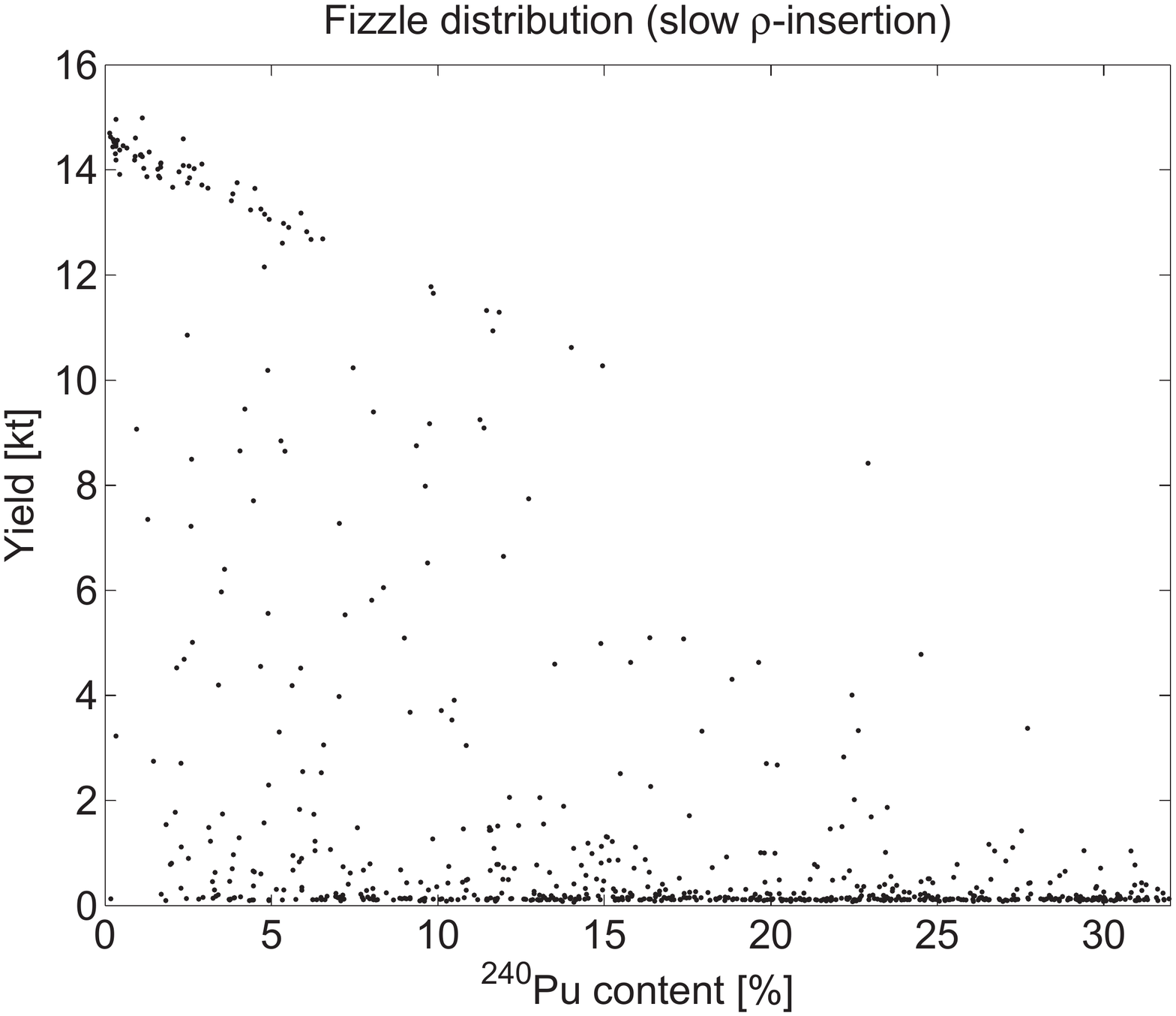}
\caption{Yield distribution for a randomly chosen $^{240}$Pu content (Trinity type device) for slow reactivity insertion.}
\label{fig_trinity_fizzle_slow}
\end{center}
\end{figure}

Fig. (\ref{fig_trinity_fizzle}) shows the yield for 700 Trinity explosions with randomly chosen neutron source terms
corresponding to a $^{240}$Pu content ranging from $0-32 \%$ (super grade to MOX grade) in the Trinity core.
The maximum compression factor reached has been randomly blurred in order to mimic different efficiencies of the high explosives
and to render the plot more legible to the eye.
For Fig. (\ref{fig_trinity_fizzle}), a compression time of $12 \rm \mu s$ was assumed from first prompt criticality until maximum compression
without predetonation \cite{Barroso}, corresponding to a velocity of the order of $1 \rm km \cdot s^{-1}$ of the surface of the
imploding plutonium core. The situation changes when the insertion time is doubled, since the fizzle probability is higher then,
as depicted in Fig. (\ref{fig_trinity_fizzle_slow}).
Note that for the simulation of the trinity device, it was assumed that the chain reaction starts at full compression
when the core is basically at rest. For fizzles, the contraction phase has to be modeled during which the core reaches first
prompt criticality where $\alpha=0$. Shortly after having reached the maximum reactivity $\rho_{max}$,
the expansion phase starts and the core eventually reaches second prompt criticality where $\alpha=0$ again. 
One should note that even when the Trinity device fizzles in the most unfavorable manner (from the war strategic view),
it still produces an energy of the order of $100 \rm kt$ or more,
enough to destroy and contaminate a city district. When the chain reaction starts at a very low positive prompt reactivity,
the core still has same time to further contract, eventually leading to a considerable energy release.

The simulations show that the fizzle yield scales
as $Y_{Fiz}^{min} \sim v_c^{3/2}$, where $v_c$ is a typical initial compression velocity of the core when reaching prompt criticality.
However, precise estimations concerning the minimal fizzle yield depend strongly on how the implosion
of the core is modeled. In Fig. (\ref{fig_trinity_fizzle_slow}), in order to facilitate the probabilistic comparison with Fig. (\ref{fig_trinity_fizzle}),
it was assumed that the homogeneously contracting core has a higher compressibility and still reaches the full compression factor $c=2.5$
when the bomb does not fizzle, although the initial contraction velocity is lower. In reality, the inserted reactivity would decrease
for a lower compression velocity, leading to a corresponding reduction of the maximum yield of the device.
The present work is not inteneded to model implosion scenarios, which belong to the harder part from the physical point of view.
In addition, the pressure in the hot phase of the fizzle explosion is no longer radiation but matter dominated
for low yields, and the limits of our model are reached. Of course, the transition to the low temperature domain where the
gas pressure $p=2E/(3V)$ becomes dominant could also be integrated in the point kinetic model.

To take it with a grain of salt, one may argue that it is no
longer relevant for the construction of a fizzle bomb with low quality plutonium to integrate an efficient neutron source in the
core, but to achieve a fast compression of the core with efficient high explosives.
Furthermore, a high $^{240}$Pu content is disadvantageous for the (fizzle) yield since the fission cross section
of $^{240}$Pu averaged over the fission spectrum is $1.349 \rm b$ \cite{JEF-2}, whereas for $^{239}$Pu one has $1.8 \rm b$.
The impact of $^{241}$Pu is relatively small.
The construction of a bomb with low quality plutonium would therefore involve high costs and comparatively low explosive yield.

\section{Conclusions}
It is a striking fact that sophisticated simulations of nuclear explosions do not necessarily lead to  better results
from a numerical point of view for many characteristic quantities than an analysis based on high school mathematics
like the one presented in this work. This is due to the complexity of the non-equilibrium thermodynamics involved
in nuclear explosions and the lack of some specific exact experimental data. The present paper is a reminiscence
of Robert Serber's lectures \cite{Serber} which were given in 1943 to new
members of the Manhattan project with the aim to explain the basic scientific facts of the wartime enterprise,
and which assembled in note form and mimeographed became the legendary
LA-1, the Los Alamos Primer, classified Secret-Limited for twenty years after the Second World War.

Despite its simplifications, the point kinetic model provides a flexible approach to the physics of the uncontrolled chain reaction.
Performing the corresponding simulations is straightforward and may help students to interpret the
extreme physical conditions that briefly occur within nuclear fission bombs. The influence of engineering
details like tampers and other material surrounding the core can be studied. Effects which are relevant, e.g., for
discussions concerning the non-proliferation of nuclear weapons, can be accessed with ease.
One-dimensional numerical simulations beyond the point kinetic model including radiative transport phenomena
for the spherically symmetric case will be presented in a forthcoming paper. 

The lesson that relatively simple 'toy model'
calculations are sufficient to get approximate estimates of the yields of nuclear detonations does not constitute a real security problem.
The main obstacle for terrorist organizations to build an atomic bomb is the acquisition and technical handling of the potentially highly radiotoxic
material with an acceptable isotopic vector needed for an efficient device. However, for developing countries, such problems
will play an increasingly smaller role in the near future. Possession of nuclear weapons may prevent states from
entering into destructive wars. Emerging decadent societies in decadent democracies, which are already in possession
of numerous nuclear weapons and which are no longer able to elect wise people as their leaders or to control their nuclear inventory,
may initiate the final countdown to zero for the third atomic bomb that will be deployed in an act of war.
In such a case the question remains whether the people are still able to apply ideas from the Age of Enlightenment,
like they are found in the Declaration of Independence, urging the people to act. 

For the sake of future generations, it is the duty of every person never to forget about and to warn of the terrible
destructive force of (thermo-)nuclear and other weapons of mass destruction.

\section{Acknowledgment}
The author wishes to thank Josef Ochsner from the Paul Scherrer Institute in Switzerland for valuable comments and
carefully reading the manuscript.

\end{document}